\documentclass[twocolumn]{jpsj2} 
\def\lsim{\mathstrut_{\displaystyle \sim}^{\displaystyle <}}
\def\gsim{\mathstrut_{\displaystyle \sim}^{\displaystyle >}}

\title{Nonlocal Excitation Spectra in 2D Doped Hubbard Model
}

\author{
Yoshiro \textsc{Kakehashi}$^{1}$\thanks{E-mail address:
yok@sci.u-ryukyu.ac.jp} and Peter \textsc{Fulde}$^{2}$\thanks{E-mail
address: fulde@mpipks-dresden.mpg.de:
to be published in J. Phys. Soc. Jpn. Vol. 76, No. 7 (2007)}
}

\inst{
$^{1}$Department of Physics and Earth Sciences,
Faculty of Science, University of Ryukyus, \\
1 Senbaru, Nishihara, Okinawa, 903-0213, Japan \\
$^{2}$Max-Planck-Institut f\"ur Physik komplexer Systeme,
N\"othnitzer Str. 38, D-01187 Dresden, Germany
}

\abst{
Single-particle excitation spectra of the two-dimensional Hubbard
model on the square lattice near half filling and at zero temperature are
investigated on the basis of the self-consistent projection operator 
method.  
The method guarantees a high accuracy of the spectra with high energy and
high momentum resolutions. It takes into account long-range intersite 
correlations as well as the strong on-site correlations. 
Effects of nonlocal excitations are clarified by comparing the results
with those of the single-site approximation.
The calculated spectra verify the quantum Monte-Carlo results 
for finite temperatures. The spectra at the Fermi level 
yield a hole-like Fermi surface in the underdoped region and 
an electron-like Fermi surface in the overdoped region.  From
a numerical analysis of the momentum dependent effective mass and
self-energy, 
it is concluded that a marginal Fermi-liquid like state
persists even at finite doping concentrations in the strongly
correlated region because a van Hove singularity is pinned
to the Fermi surface.  It is also found that a kink structure appears
in the quasiparticle energy band in the same region.  
The kink is shown to be caused by a mixing between the quasiparticle
band and an excitation band with strong short-range antiferromagnetic
correlations. These results suggest an explanation for some of 
the unusual properties
of the normal state in high-$T_{\rm c}$ cuprates. 
}

\kword{quasiparticle excitations, two-dimensional Hubbard model, 
momentum-dependent self-energy, marginal Fermi liquid, ARPES, 
kink, cuprate, LSCO}

\begin{document}
\maketitle

\section{Introduction} 
The two-dimensional (2D) Hubbard model has been widely investigated
in the past two decades because it has been the simplest 
model for describing the high-temperature superconductivity in layered 
Cu-based perovskites~\cite{dagotto94,imada88,maier05}.  
The Cu 3d$_{x^2 - y^2}$ orbitals on a square lattice hybridize with the O 
2p${}_{x}$ and 2p${}_{y}$ orbitals on the CuO${}_{2}$ plane and form an 
antibonding band near the Fermi level. There is a strong on-site Coulomb 
repulsion of electrons in hybridized Cu 3d orbitals with different
spins, which leads to the 2D Hubbard 
model~\cite{anderson97,schuttler92,simon93}.

The 2D Hubbard model at half-filling with nearest-neighbor electron hopping is
an antiferromagnetic insulator with the Ne\'{e}l temperature 
$T_{\rm N}=0$ according to the Mermin-Wagner theorem~\cite{mermin}.
When holes are doped, charge fluctuations counteract antiferromagnetism and a
metallic state should be realized. Rigorous results are however scarce
especially for intermediate Coulomb interaction strengths, and therefore the
model is of current interest in condensed matter physics.
 
Numerical techniques applied to the doped regime such as the Lanczos
method~\cite{dagotto94} and Quantum Monte-Carlo
(QMC)~\cite{bulut94,preuss94,grober00}  have verified the formation of a gap at
half-filling and clarified global features of excitation spectra from the
insulator to a metal. The antiferromagnetic correlations
have been shown to reduce rapidly with increasing hole
concentration~\cite{grober00,furukawa}.   More recent calculations based on the
dynamical cluster approximation~\cite{jarrell01-1} verified a hole-like Fermi
surface in the underdoped region and an electron-like Fermi surface in the overdoped
region~\cite{maier02}. The calculations also suggested the possibilities of a
pseudogap state as well as the occurrence of
superconductivity~\cite{jarrell01-2}. 

The methods mentioned above are based on a cluster approach.
The approach treats a finite cluster explicitly 
by assuming a boundary condition, or implicitly by embedding it in an
effective medium. Since it is numerically not possible to treat a large
cluster, finite size effects are inevitable in these methods. In fact, the
methods do not yield high momentum and energy resolutions for excitation
spectra, and do not describe the long-range intersite correlations due to
finite cluster size~\cite{kake04-1}. 
It is difficult to achieve a momentum
resolution better than 0.005 $\dot{\rm A}^{-1}$ which was recently obtained 
in angle resolved photoemission spectroscopy (ARPES)\cite{johnson}.  
This difficulty becomes serious for the investigations of the kink
structure of the quasiparticle band found in the cuprates.  
Therefore one needs different approaches to study a more detailed structure of 
spectra in the 2D Hubbard model.

In this paper we investigate the single-particle excitation 
spectra and related properties of the 2D Hubbard model on the square lattice by
making use of the self-consistent projection operator method
(SCPM)~\cite{kake04-2}. The SCPM is an extension of the projection
operator method which uses the coherent potential approximation 
(PM-CPA)~\cite{kake04-3}, 
when determining the momentum dependent self-energy. The SCPM treats directly
the retarded Green function, describes the local on-site correlations by using
an effective medium, and takes into account the intersite correlations up to
infinity by using an incremental cluster expansion~\cite{stoll90,fulde02}.  
Thus the method yields high momentum and energy resolutions, and allows us to
determine the detailed structure of nonlocal excitations in both momentum and
energy spaces.   
Using the SCPM, we will reveal the nonlocal effects on the excitation 
spectra of the 2D
Hubbard model and its momentum dependent properties at zero temperature,
which could not be clarified by means of the other methods.

The high-$T_{\rm c}$ cuprates have unusual properties in the normal
state; the resistivity shows a $T$-linear temperature dependence and 
a temperature independent term appears in the nuclear relaxation 
rate~\cite{norman03}. These features were first explained by a phenomenological
marginal Fermi liquid (MFL) scenario~\cite{varma89}. The MFL is defined by
electrons whose imaginary part of the self-energy near the Fermi surface, ${\rm Im} \Sigma(\omega, T)$ is proportional to $|\omega|$ instead of $\omega^{2}$
where $\omega \ (> T)$ is an excitation energy and $T$ is the temperature.
Accordingly, Re $\Sigma(\omega, T) \propto \omega {\rm ln} |\omega|$.
In spite of successes of the theory, the MFL has remained puzzling because no
microscopic theory could be provided for it. We show in the present paper that
a MFL like behavior appears in the underdoped region, which may be considered
as a form of justification of the phenomenological MFL theory.  

Another interesting point of our results is that a kink
structure in the quasiparticle band appears in the underdoped region.
Recent ARPES experiments have revealed a kink in the
quasiparticle band of high-$T_{\rm c}$ cuprates in both the normal
and superconducting state~\cite{bodanov,lanzara,cuk}. The typical kink energy
at which the quasiparticle velocity changes is 70 meV. Previous theories for
the kink structure assumed either a coupling of electrons to a magnetic
resonance mode found in inelastic neutron scattering
experiments~\cite{johnson,eschrig}  or a coupling to a longitudinal optical
phonon mode~\cite{cuk,schach}. These theories do not take into account other
electron correlations which are an essential ingredient of the high-$T_{\rm c}$
cuprates.  We demonstrate on the basis of the SCPM that the kink structure can
be explained by electron correlations without introducing lattice degrees of
freedom. This provides a new possible mechanism for the observed kink structure
in cuprates.  Preliminary results on these topics have been published
recently~\cite{kake05-1,kake05-2}.  
It should be noted that the validity of numerical calculations on these 
problems strongly depends on the momentum and energy resolutions.
We therefore have recalculated the spectra increasing the number of mesh 
from $80 \times 80$ to $160 \times 160$ in the first Brillouin zone and 
the energy mesh by 
a factor of two.  On the basis of the new results of calculations 
we will discuss in the present paper the MFL and the kink structure 
in more details.

The paper is organized as follows. In the next section, we review briefly the
SCPM to the Hubbard model. In Sec. III we present our results of calculations
for the excitation spectra.  We limit ourselves here to excitation spectra in
the normal state at zero temperature. In Sec. III A we discuss the
characteristics of the excitation spectra at half-filling under the assumption
of a paramagnetic effective medium. In Sec. III B, we treat the doped
case.  
We present the excitation spectra, integrated density of states (DOS), Fermi
surface, momentum distribution, and the momentum dependent effective
mass.  
We clarify the nonlocal effects of electron correlations on these
quantities by comparing them with those of the single-site approximation
(SSA) and other numerical results.  
The MFL-like behavior which we find in the underdoped
region is discussed  separately in Sec. III C. The MFL features are shown to be
caused by a pinning of the van Hove singularity to the Fermi surface under
doping.  In Sec. III D, we discuss the kink structure of the quasiparticle
band, which is obtained in the underdoped region. The results of our
calculations are summarized in Sec IV, where also the unusual behavior of
cuprates in the normal state is discussed.

\section{Self-consistent projection operator method}

We consider a Hubbard model on a square lattice with an atomic level 
$\epsilon_{0}$, a nearest-neighbor transfer integral $t_{ij}\, (=t)$, 
and an on-site Coulomb interaction parameter $U$.
\begin{eqnarray}
H = \sum_{i,\sigma} \, (\epsilon_{0} - \mu) \, n_{i \sigma}
+ \sum_{i, j, \sigma} t_{i j} \, a_{i \sigma}^{\dagger} 
a_{j \sigma} 
+ U \sum_{i} \, n_{i \uparrow} n_{i \downarrow} \ .
\label{hub}
\end{eqnarray}
Here the chemical potential $\mu$ has been added to the
Hamiltonian. Furthermore $a_{i \sigma}^{\dagger}$ ($a_{i \sigma}$) is the
creation (annihilation) operator for an electron with spin $\sigma$ on site $i$
and $n_{i \sigma}=a_{i \sigma}^{\dagger}a_{i \sigma}$.

The single-particle excitation spectrum is obtained from the poles of the
retarded Green function given by
\begin{eqnarray}
G_{k\sigma}(z) = \frac{1}{z - \epsilon_{k\sigma} - 
\Lambda_{k\sigma}(z)} \ .
\label{gk}
\end{eqnarray}
Here $z=\omega+i\delta$ with $\delta$ being an infinitesimal positive
number,  $\epsilon_{k\sigma}$ is the Hartree-Fock energy given by 
$\epsilon_{k\sigma} = \epsilon_{0} + \epsilon_{k} - \mu + U \langle
n_{i-\sigma} \rangle$ where $\langle n_{i\sigma} \rangle$  and $\epsilon_{k}$ denote
the average number of electrons with spin $\sigma$ on site $i$ and the dispersive
part of the energy band for noninteracting electrons, respectively.

The self-energy $\Lambda_{k\sigma}(z)$ is given by a Fourier transform
of the memory function $M_{ij\sigma}(z)$ as
\begin{eqnarray}
\Lambda_{k\sigma}(z) = U^{2} 
\sum_{j} M_{j0\sigma}(z) 
\exp (i\mbox{\boldmath$k$}\cdot\mbox{\boldmath$R$}_{j}) \ .
\label{memk}
\end{eqnarray}
In the projection operator method~\cite{fulde,lukas82,kishore87}, 
the memory function is given by
\begin{eqnarray}
M_{ij\sigma}(z) = 
\left( A^{\dagger}_{i \sigma} \  {\Bigl |} \ (z-\overline{L})^{-1} \,
A^{\dagger}_{j \sigma} \right) \ .
\label{rmem}
\end{eqnarray}
Here 
$A^{\dagger}_{i \sigma}=a^{\dagger}_{i \sigma} ( n_{i-\sigma} -
\langle n_{i-\sigma} \rangle)$. The inner product of operators A and B is
defined by $(A|B)=\langle [A^{+},B]_{+} \rangle$ where the expectation 
value is taken
with respect to the ground state of the system. The dynamics of the electronic
system is described by a superoperator $L$. It acts on other operators $A$
according to $LA = [H, A]_-$. Furthermore $\overline{L} = QLQ$ where
the projector $Q$ is given by $Q = 1-P$ and $P = \sum_{i \sigma}
|a^{\dagger}_{i \sigma}) (a^{\dagger}_{i \sigma}|$. Thus $Q$ eliminates
$|a^{\dagger}_{i \sigma})$ from further considerations.

In the SCPM~\cite{kake04-2} we introduce an effective Liouville operator 
\begin{eqnarray}
\tilde{L}(z) A = [\tilde{H}(z),A]_{-} \ ,
\label{leff}
\end{eqnarray}
\begin{eqnarray}
\tilde{H}(z) = H_{0} + \sum_{i\sigma} 
\tilde{\Sigma}_{\sigma}(z) \, n_{i \sigma} \ .
\label{heff}
\end{eqnarray}
Here $H_{0}$ is the Hartree-Fock Hamiltonian, and $\tilde{\Sigma}_{\sigma}(z)$
defines an energy-dependent effective medium and is called the coherent
potential.  The Liouvillean $L$ is divided into a coherent part $\tilde{L}(z)$
and an interaction part $L_{\rm I}(z)$.  By making use of the multiple
scattering theory, we can derive an incremental cluster expansion for the
memory functions: 
\begin{eqnarray}
M_{ii \sigma}(z) &=&
M^{(i)}_{ii \sigma}(z)
+ \sum_{l \neq i} \Delta M^{(il)}_{ii \sigma}(z) 
\hspace{20mm} \nonumber \\
&+& \, \frac{1}{2} \, {\sum_{l \neq i}} \,{\sum_{m \neq i,l}} \, 
\Delta M^{(ilm)}_{ii \sigma}(z) + \cdots \ ,
\label{icrii}
\end{eqnarray}
\begin{eqnarray}
M_{ij \sigma}(z) &=&
M^{(ij)}_{ij \sigma}(z) + 
\sum_{l \neq i,j} \Delta M^{(ijl)}_{ij \sigma}(z)
\hspace{20mm} \nonumber \\
&+& \, \frac{1}{2} \, \sum_{l \neq i,j} \sum_{m \neq i,j,l} \, 
\Delta M^{(ijlm)}_{ij \sigma}(z) + \cdots \ ,
\label{icrij}
\end{eqnarray}
with
\begin{eqnarray}
\Delta M^{(il)}_{ii\sigma}(z) = 
M^{(il)}_{ii\sigma}(z) - M^{(i)}_{ii\sigma}(z) \ ,
\label{icr2ii}
\end{eqnarray}
\begin{eqnarray}
\Delta M^{(ilm)}_{ii\sigma}(z) = 
M^{(ilm)}_{ii\sigma}(z) 
\hspace{32mm} \nonumber \\
- \Delta M^{(il)}_{ii\sigma}(z) 
- \Delta M^{(im)}_{ii\sigma}(z) 
- M^{(i)}_{ii\sigma}(z)\ ,
\label{icr3ii}
\end{eqnarray}
\begin{eqnarray}
\Delta M^{(ijl)}_{ij\sigma}(z) = 
M^{(ijl)}_{ij\sigma}(z) - M^{(ij)}_{ij\sigma}(z) \ ,
\hspace{15mm} 
\label{icr2ij}
\end{eqnarray}
\begin{eqnarray}
\Delta M^{(ijlm)}_{ij\sigma}(z) = 
M^{(ijlm)}_{ij\sigma}(z) 
\hspace{30mm} \nonumber \\
- \Delta M^{(ijl)}_{ij\sigma}(z) 
- \Delta M^{(ijm)}_{ij\sigma}(z) - M^{(ij)}_{ij\sigma}(z)\ .
\label{icr3ij}
\end{eqnarray}

In the above expressions, $M^{(c)}_{ij\sigma}(z)$ are cluster memory
functions defined by
\begin{eqnarray}
M^{(c)}_{ij\sigma}(z) =  
\Big( A^{\dagger}_{i\sigma} \  {\Bigl |} (z-\overline{L}^{(c)}(z))^{-1} \,
A^{\dagger}_{j\sigma} \Big) \ ,
\label{gpij}
\end{eqnarray}
\begin{eqnarray}
L^{(c)}(z) A = 
\hspace{60mm} \nonumber \\
\Bigl[\tilde{H}(z) -\sum^{N_{c}}_{i \in c} 
(\sum_{\sigma} \tilde{\Sigma}_{\sigma}(z) 
n_{i\sigma} -
U \delta n_{i\uparrow} \delta n_{i\downarrow}) \, , A \Bigr]_{-} \ .
\label{li}
\end{eqnarray}
Here $\overline{L}^{(c)}(z)=QL^{(c)}(z)Q$, and $\delta n_{i\sigma} =
n_{i\sigma} - \langle n_{i\sigma} \rangle$. $L^{(c)}(z)$ is the Liouvillean for
the cluster $c$ which is embedded in an effective medium
$\tilde{\Sigma}_{\sigma}(z)$, and $N_{c}$ denotes the number of atoms in the
cluster. It should be noted that we take into account all the 'clusters' obtained
by choosing any $N_{c}$ sites from the 2D square lattice points.  This
is quite different from the usual cluster approaches in which only one compact
cluster of sites with a given form is considered. It should also be noted that
when the off-diagonal memory function (eq.(\ref{icrij})) is neglected and only
the single-site term (SSA) is taken into account in eq. (\ref{icrii}), 
the SCPM reduces to the PM-CPA, which is equivalent to the many-body 
CPA~\cite{hirooka}, 
the dynamical CPA~\cite{kake92}, and the dynamical mean-field
theory\cite{georges96,kake022}.   

We use here the memory functions obtained by the renormalized
perturbation scheme~\cite{kake04-2}:
\begin{eqnarray}
M^{(c)}_{ij\sigma}(z) = 
\Big[ \mbox{\boldmath$M$}^{(c)}_{0} \cdot 
(1-\mbox{\boldmath$L$}^{(c)}_{I} \cdot 
\mbox{\boldmath$M$}^{(c)}_{0})^{-1} \Big]_{ij\sigma} \ .
\label{memcij}
\end{eqnarray}
The diagonal matrix $\overline{\mbox{\boldmath$L$}}^{(c)}_{I}$
is defined by~\cite{comm1}
\begin{eqnarray}
\overline{\mbox{\boldmath$L$}}^{(c)}_{I} = 
[\overline{L}^{(i)}_{I\sigma}, \overline{L}^{(j)}_{I\sigma}, 
\cdots, \overline{L}^{(l)}_{I\sigma}] \ ,
\label{lic}
\end{eqnarray}
\begin{eqnarray}
\overline{L}^{(i)}_{I\sigma}(z) =
\frac{\displaystyle U (1- 2 \langle n_{i -\sigma} \rangle)}
{\displaystyle \langle n_{i -\sigma} \rangle 
(1 - \langle n_{i -\sigma} \rangle)} \ .
\label{liint}
\end{eqnarray}
The screened memory function $\hat{\mbox{\boldmath$M$}}^{(c)}_{0}$ 
in eq. (\ref{memcij}) is a
$N_{c} \times N_{c}$ matrix given by
\begin{eqnarray}
M^{(c)}_{0ij\sigma}(z) = 
\hspace{62mm} \nonumber \\
A_{ij\sigma} \!\!\!
\int \frac{\displaystyle d\epsilon d\epsilon^{\prime} 
d\epsilon^{\prime\prime} 
\tilde{\rho}^{(c)}_{ij\sigma}(\epsilon)
\tilde{\rho}^{(c)}_{ij-\sigma}(\epsilon^{\prime})
\tilde{\rho}^{(c)}_{ji-\sigma}(\epsilon^{\prime\prime})
\chi(\epsilon, \epsilon^{\prime},
\epsilon^{\prime\prime})}
{\displaystyle z  -
\epsilon - \epsilon^{\prime} + \epsilon^{\prime\prime}}
, 
\label{lrpt}
\end{eqnarray}
\begin{eqnarray}
A_{ij\sigma} = \frac{\displaystyle \langle n_{i -\sigma} \rangle 
(1 - \langle n_{i -\sigma} \rangle)}{\langle n_{i -\sigma} \rangle_{0} 
(1 - \langle n_{i -\sigma} \rangle_{0})}\delta_{ij} + 1 - \delta_{ij} \ ,
\label{aij}
\end{eqnarray}
\begin{eqnarray}
\chi(\epsilon, \epsilon^{\prime},\epsilon^{\prime\prime}) =
f(-\epsilon)f(-\epsilon^{\prime})f(\epsilon^{\prime\prime})
+ f(\epsilon)f(\epsilon^{\prime})f(-\epsilon^{\prime\prime}) \ .
\label{chi}
\end{eqnarray}

The memory function $\hat{M}^{(c)}_{0ij\sigma}$ consists of that of the
second-order perturbation~\cite{viro90,schweizer} theory and of a prefactor
$A_{ij\sigma}$ ensuring the correct second moment in the moment
expansion. Furthermore $\langle n_{i \sigma} \rangle_{0}$ is the average
electron number defined by $\langle n_{i\sigma} \rangle_{0} = \int d\omega
f(\omega)\tilde{\rho}^{(c)}_{ii\sigma}(\omega)$ where $f(\epsilon)$ is the
Fermi distribution function. The density of states
$\rho^{(c)}_{ij\sigma}(\epsilon)$ is defined by 
\begin{eqnarray}
\tilde{\rho}^{(c)}_{ij\sigma}(\epsilon) = 
-\frac{1}{\pi} \, {\rm Im} \, [(\tilde{\mbox{\boldmath$F$}}(z)^{-1}
+ \tilde{\mbox{\boldmath$\Sigma$}}^{(c)}(z)
)^{-1}]_{ij\sigma} \ ,
\label{rhocav}
\end{eqnarray}
\begin{eqnarray}
(\tilde{\mbox{\boldmath$F$}}(z))_{ij\sigma} = 
\int \frac{\rho_{ij}(\epsilon) \, d\epsilon}
{z - \epsilon_{\sigma} - \tilde{\Sigma}_{\sigma}(z)-\epsilon} \ ,
\label{ftl}
\end{eqnarray}
where $\epsilon_{\sigma}=\epsilon_{0}- \mu + U \langle n_{i-\sigma} \rangle$, and
$\rho_{ij}(\epsilon)$ is the density of states for a noninteracting system,
\begin{eqnarray}
\rho_{ij}(\epsilon) = \frac{1}{N} \sum_{\mbox{\boldmath$k$}} 
\delta(\epsilon-\epsilon_{k})
\exp [ -i \mbox{\boldmath$k$} \cdot 
(\mbox{\boldmath$R$}_{i}-\mbox{\boldmath$R$}_{j})] \ .
\label{rhoij}
\end{eqnarray}
The coherent potential introduced in eq. (\ref{heff}) is determined self-consistently 
from the CPA equation, i.e.,
\begin{eqnarray}
\tilde{\Sigma}_{\sigma}(z)=\Lambda_{ii\sigma}(z)=
N^{-1}\sum_{k}\Lambda_{k\sigma}(z) \ .
\label{cpa}
\end{eqnarray}

Equations (\ref{memk}), (\ref{icrii}), (\ref{icrij}), (\ref{memcij}), and
(\ref{cpa}) form a self-consistent set of equations from which we can obtain
the effective medium $\tilde{\Sigma}_{\sigma}(z)$.  The nonlocal excitations
are then obtained from eq. (\ref{gk}).  We have solved the self-consistent
equations at zero temperature within the two-site approximation in which we take
into account the first two terms of the right hand side of eq. (\ref{icrii})
and the first term in eq. (\ref{icrij}). 

\section{Numerical results}
\subsection{Nonlocal excitations of the half-filled band}

The 2D Hubbard model is considered to be an antiferromagnetic insulator
with the Ne\'{e}l temperature $T_{\rm N}=0$ K at half filling~\cite{mermin}.
In the effective medium approach there are two solutions found, a nonmagnetic
solution and an antiferromagnetic (AF) one. The latter describes the system
at half filling. But in the SCPM an AF effective medium overestimates
antiferromagnetic correlations at finite doping concentrations because the
medium eliminates spin fluctuations. Therefore the SCPM does not describe
properly the AF at $T = 0$.  Here we study the properties of
excitations in the presence of the nonmagnetic medium, because the nonmagnetic
solution  at half-filling is smoothly connected to that at finite doping
concentration.  In the following we present results for the nonlocal
excitations in the nonmagnetic medium at half-filling. 
\begin{figure}
\includegraphics[width=9.5cm]{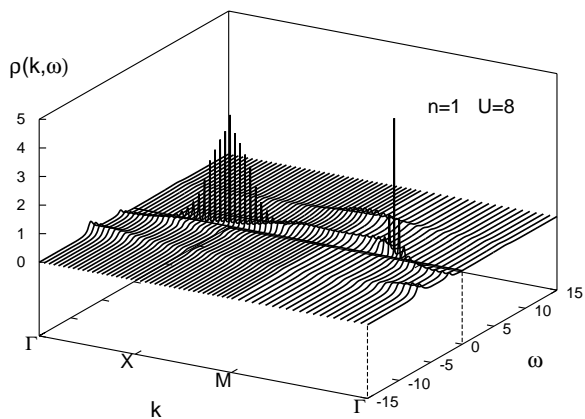}%
\caption{\label{dosk1}
Single-particle excitation spectrum along the high symmetry lines for
half filling ($n=1$) and $U=8$.  The Fermi level is indicated by a bold line.
Here and in the followings, the energy is measured in unit of $|t|$.
Note that $\Gamma=(0,0), X=(\pi,0), M=(\pi,\pi)$ are in unit of the lattice
 constant. 
}
\end{figure}
\begin{figure}
\includegraphics[width=9.5cm]{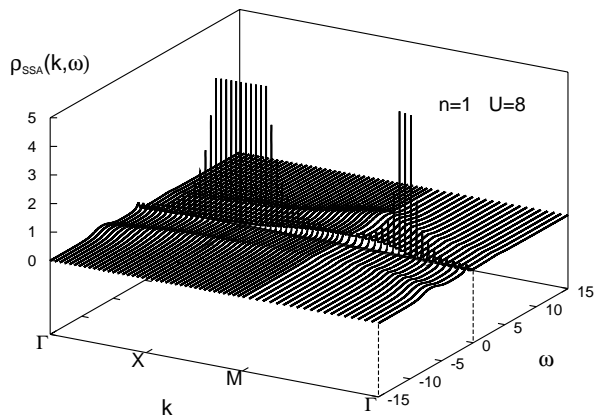}%
\caption{\label{dosk0}
Single-particle excitation spectra in the single-site approximation
 (SSA).  Parameters are the same as in Fig.~\ref{dosk1}.
}
\end{figure}

For the numerical calculations, we adopted a
$160 \times 160$ (or $80 \times 80$) mesh in the first Brillouin zone, and
obtained the momentum dependent self-energy (\ref{memk}) by taking into account
the off-diagonal memory functions up to the 50th (!) nearest neighbors.  The
latter are calculated from eq. (\ref{memcij}). The screened memory functions
$M^{(c)}_{0ij\sigma}(z)$ in eq. (\ref{memcij}) are calculated by means of 
the Laplace transformation method~\cite{schweizer}. In order to avoid
singularities in the Green functions, we used a complex energy $z=\omega +
i\delta$ with $\delta = 0.025$ (or $\delta = 0.05$) in unit of $|t|$.  

Figure \ref{dosk1} shows the excitation spectrum along the high symmetry line
at $U=8$. The spectrum is characterized by incoherent Mott-Hubbard excitations,
i.e., a lower Hubbard band (LHB) around the $\Gamma$ point and an upper Hubbard
band (UHB) around the M point. In addition there is the quasiparticle band near
the Fermi energy ($\omega = 0$). The nonlocal excitation spectrum is compared
with that in the single-site approximations (see Fig.~\ref{dosk0}). We find
that nonlocal correlations reduce the amplitude of the quasiparticle peak near
the Fermi level, and increase its band width by a factor of two as
compared with the SSA for an
intermediate strength of the repulsion $U$. Alternatively, the electron
correlations in the SSA overestimate the band narrowing in the quasiparticle state.
When intersite correlations are additionally included, the electrons recover
parts of the original degrees of freedom for their motion. The result is a band
broadening but the Coulomb repulsion energy remains optimized.

Nonlocal excitations also enhance the amplitude of the LHB around the $\Gamma$
point as well as that of the UHB around the M point. Similar global features of
these nonlocal excitation spectra are also found in the 3D Hubbard model on a
simple-cubic lattice~\cite{kake04-2}. Another characteristic of the spectra is
pronounced flat-band excitations at $\omega \approx \pm 2.0$ due to
antiferromagnetic correlations. These flat bands are precursors of an
antiferromagnetic insulator~\cite{jarrell01-1}. They should develop further
when higher-order cluster correlations are taken into account by going beyond
the two-site approximation. 
\begin{figure}
\includegraphics[width=9cm]{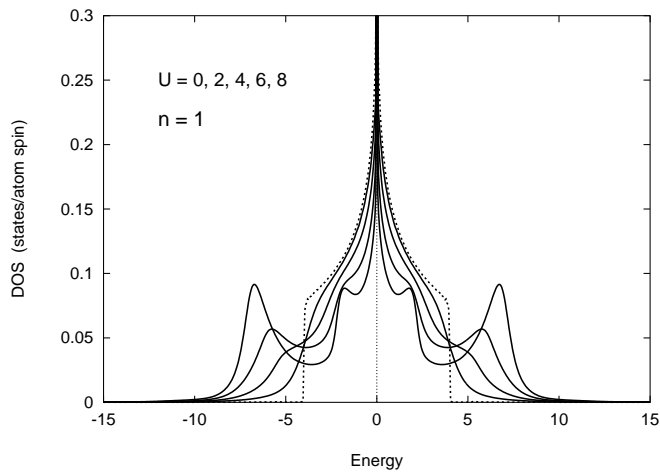}%
\caption{\label{dosn1u}
Total densities of states (DOS) at half filling as function of $U$.
Noninteracting DOS is shown by dotted line.
}
\end{figure}

The total density of states (DOS) as a function of interaction $U$
is presented in Fig.~\ref{dosn1u}. A characteristic feature of the DOS is a
logarithmic divergence at $\omega = 0$, which is due to a van Hove singularity
at the X point. 
For a small Coulomb interaction, we can compare our results with those
of the perturbation theory~\cite{zlatic00,shinkai05}.
We find a good agreement between the presend DOS and that obtained by
the 4th order perturbation theory for $U=2$.  For $U=4$, the shoulder
position is $|\omega|=5.0$ in the present calculation, while it is
$|\omega|=4.5$ in the 4th order perturbation.  These results verify 
the validity of our results for the regime of Coulomb interaction
$U \lsim 4$, in which the 4th order perturbation is believed to be
valid.  
The quasiparticle band in the low energy region becomes narrow
when the interaction $U$ is increased. The Mott-Hubbard incoherent bands on
the other hand develop in the high energy regions. Furthermore, the nonlocal
antiferromagnetic correlations create subbands at $|\omega| \approx 1.8$ when
$U \gsim 6$.  When $U \gsim 10$, we find within the two-site approximation a
negative spectral density in a small region of momentum space. Therefore we
have to include higher-order cluster correlations in order to discuss the
spectra beyond $U \gsim 10$ within the present scheme. In the following we
limit our discussions to the metallic region $U \lsim 10$.
\begin{figure}
\includegraphics[width=8.5cm]{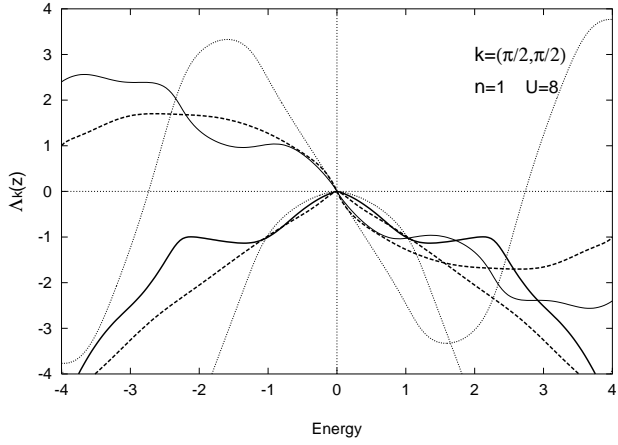}%
\caption{\label{skf1}
Self-energies at the Fermi momentum ($\pi/2,\pi/2$) in various 
approximations; Two-site approximation (solid curves), second-order
perturbation (thick dashed curves), single-site approximation (thin
dashed curves). Note that the real part is antisymmetric, and the imaginary
part is symmetric in energy. 
 }
\end{figure}

A characteristic feature of the 2D Hubbard model at half filling is that of
marginal Fermi liquid.  The MFL is defined by Re $\Lambda_{k_{F}}(z) \propto
\omega \, {\rm ln} \, |\omega|$ and Im $\Lambda_{k_{F}}(z) \propto |\omega|$
for small $|\omega|$ and for the Fermi wave number $k_F$. In the weakly correlated 
regime such behavior is found by 2nd-order perturbation
theory~\cite{viro90,schweizer}. For the intermediate Coulomb interaction
strength, we have calculated the self-energy at the Fermi surface.  An example
is shown in  Fig. \ref{skf1}. The MFL behavior seems to be for $k$ present even
for strong Coulomb interaction $U$, although numerically the linear dependence
of Im $\Lambda_{k_{F}}(z)$ on $|\omega|$ is not so accurate.
\begin{figure}
  \includegraphics[width=9.5cm]{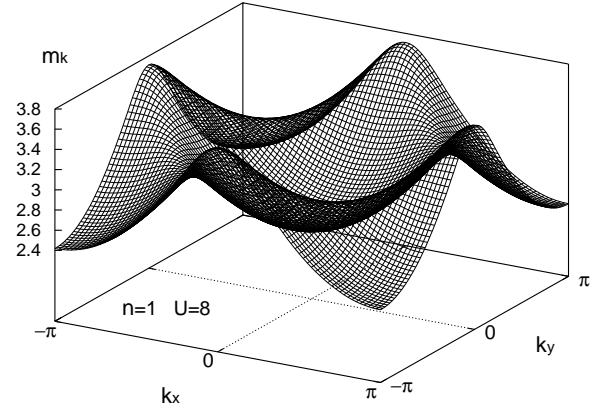}%
  \caption{\label{mk1}
Momentum dependent effective mass at half filling, calculated by
numerical differentiation by using an energy interval of $\delta\omega=0.05$.
} 
\end{figure}

We have calculated the momentum dependent effective mass $m_{k} = (1- \partial
{\rm Re} \Lambda_{k}(z)/\partial \omega)_{\omega=0}$ using numerical
differentiation. As presented in Fig.~\ref{mk1}, the calculated effective mass
shows a strong momentum dependence, having a minimum at the M point
($=(\pi,\pi)$),  and a maximum at the X point ($=(\pi,0)$). Along the Fermi
surface $\cos k_{x} + \cos k_{y} = 0$, $m_{k}$ has a minimum at $(\pi/2,\pi/2)$
and a maximum at $(\pi,0)$. This suggests the appearance of the 'Fermi
arc'~\cite{yoshida} when holes are doped.  

Because ${\rm Re} \Lambda_{k}(z) \propto \omega {\rm ln} |\omega|$ for small 
$|\omega|$ in the MFL, it is $m_{k} -1 \propto - {\rm ln} \delta\omega$ for
small energy steps $\delta\omega$ when a numerical differentiation is done.
Therefore, the simple numerical method for the calculation of $k$-dependent
effective mass breaks down in the case of the MFL.  
\begin{figure}
\includegraphics[width=9.5cm]{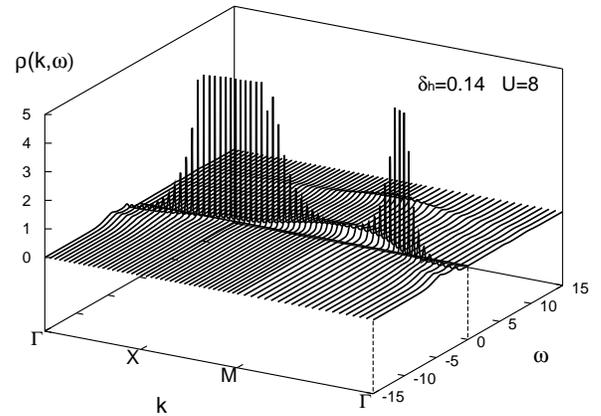}%
\caption{\label{dosk86}
Excitation spectrum along the high symmetry lines for doping concentration
$\delta_{h}=0.14$.
}
\end{figure}
We have verified that the effective mass at the Fermi surface 
in Fig.~\ref{mk1} 
increases logarithmically with decreasing $\delta \omega$, i.e., $m_{k}$ along
the Fermi surface is expected to show a logarithmic divergence. We will return
to this problem in Sec. III C.

\subsection{Nonlocal excitations in the doped Hubbard model}

When holes are doped into a half-filled 2D Hubbard system, 
quantum spin and charge fluctuations are enhanced due to a motion of
holes, and thus the system is expected to lose the long range magnetic
order.  In the very weak Coulomb interaction limit, it is suggested by
using the Hartree-Fock approximation~\cite{schulz90} that an
incommensulate antiferromagnetic order can remain even at finite doping
concentration, and the insulating state can survive under the long range
magnetic order. A problem of the theory is that a finite value of
$T_{\rm N}$ at half-filling contradicts with the Mermin-Wagner 
theorem~\cite{mermin}.  This suggests that the quantum fluctuations
are dominant in the 2D system so that the long range magnetic order
is suppressed at finite doping concentration.  Although there are no
rigorous results on this problem we assume here in our numerical
calculations the nonmagnetic state even for an infinitesimal doping 
concentration.  

A typical excitation spectrum is shown in Fig.~\ref{dosk86} for 
a doping concentration of $\delta_{h} = 0.14$. 
As seen from the figure, the upper Hubbard band moves
upwards. The lower Hubbard band has become weaker and merges with the
quasiparticle band near the $\Gamma$ point. The excitations due to 
short-range AF correlations, which are located at $|\omega| \approx 1.8$ 
in Fig.~\ref{dosk1}, have disappeared.  
Consequently, the spectral weight of quasiparticle peak is enhanced.   
\begin{figure}
\includegraphics[width=9cm]{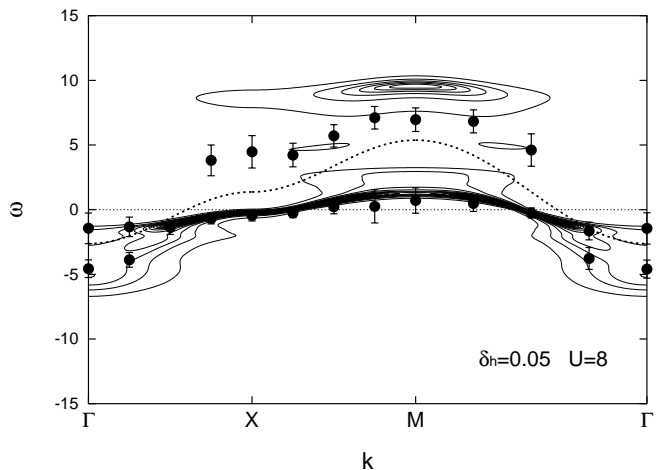}%
\caption{\label{dctr95}
Contour map of the excitations at $\delta_{h}=0.05$. Closed circles with error
bars are the QMC results~\cite{grober00} at $T=0.33$. The dashed curves show
the Hartree-Fock quasiparticle dispersion. 
}
\end{figure}

\begin{figure}
\includegraphics[width=9cm]{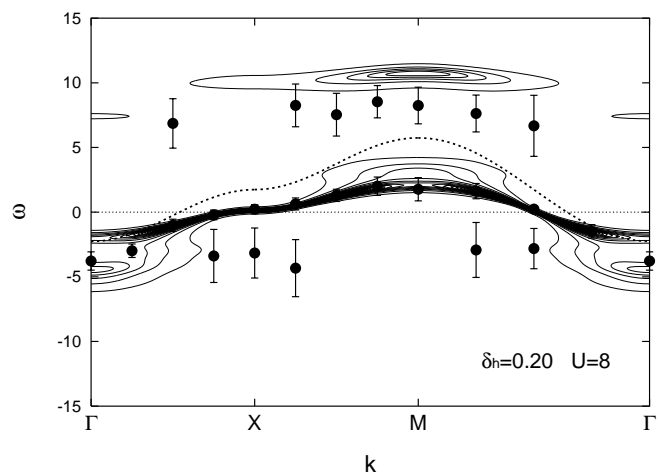}%
\caption{\label{dctr80}
Contour map of excitation spectra at $\delta_{h}=0.20$.
}
\end{figure}

We have examined the doping dependence of the excitation spectra. At 5 \%
doping, the LHB is nearly destroyed as shown in Fig. \ref{dctr95}.  The
flat bands at $|\omega| \sim 2$ around the $\Gamma$ and M point, which are due
to short-range antiferromagnetic correlations,  have been weakened. The
quasiparticle band on the other hand is well developed. Accordingly, the flat
quasiparticle band around the X point sinks below the Fermi level.
\begin{figure}
\includegraphics[width=9cm]{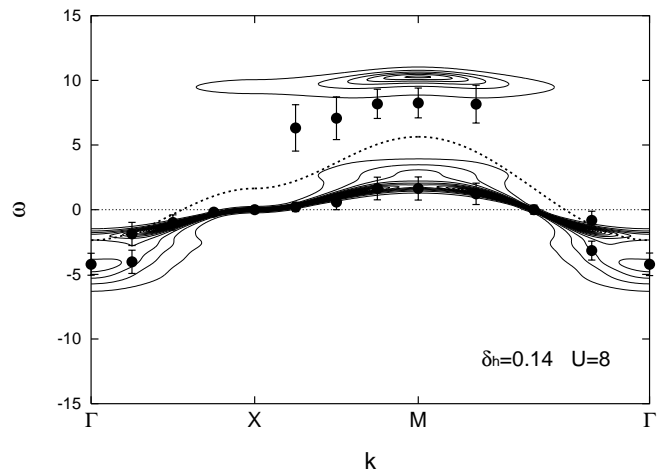}%
\caption{\label{dctr86}
Contour map of the excitations at $\delta_{h}=0.14$.
}
\end{figure}
In the optimum doped region (see Fig.~\ref{dctr86}), the quasiparticle band
is further pronounced.  But its width is broadened by 26 \% as
compared with 5 \% doping.  Note that due to the increase of
holes the flat band around the X point is at the Fermi level. 
In the overdoped region (see Fig.~\ref{dctr80}), the quasiparticle band
gains further weight and its width increases by 10 \% as compared with 14 \%
doping.  The flat band around the X point is now above the Fermi level.

We have compared our results at zero temperature with those of the QMC at
finite temperatures~\cite{grober00}. As shown by closed circles in
Figs.~\ref{dctr95}, \ref{dctr86}, and \ref{dctr80}, the present results for the
nonlocal excitations are consistent with the QMC results for the underdoped
region as well as for the overdoped one. Especially, the calculated
quasiparticle bands show quantitative agreement in both cases. In the QMC
calculations for 20 \% doping concentration, weak excitations are found at
$\omega \approx -3.0$ around the X and $(\pi/2,\pi/2)$ points as shown in
Fig.~\ref{dctr80}. Although not visible in Fig.~\ref{dctr80}, we have verified
that the same excitations do appear in our calculations. For example, we find a
small peak of height 0.03 at the X point $(\pi,0)$ and energy $\omega=-2.9$, in
agreement with the QMC results. We have also investigated the spectra in the SSA. 
The quasiparticle weight in the SSA is underestimated at the $\Gamma$ point and
is overestimated at the M point. The LHB (UHB) in the SSA is correspondingly
excessively enhanced (weakened). 
\begin{figure}
\includegraphics[width=9cm]{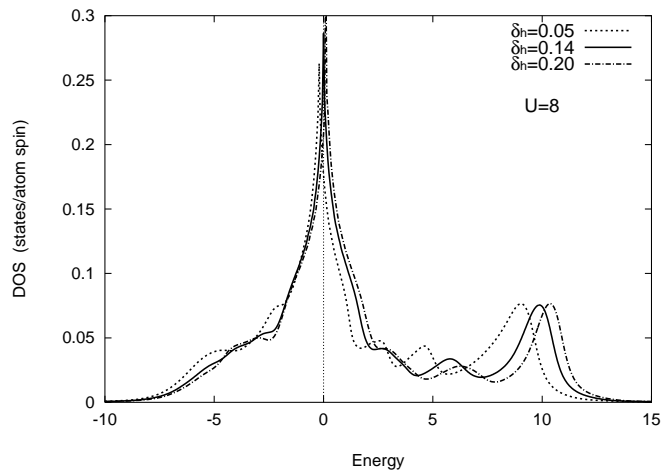}%
\caption{\label{dosun}
Total DOS for various doping concentrations
}
\end{figure}

The integrated DOS's are shown in Fig.~\ref{dosun} for different hole dopings
$\delta_{h}=$ 0.05, 0.14, and 0.20.  We find that the weight of the LHB is
rather small for these doping concentrations. The upper Hubbard band simply
shifts to the higher energy region with increasing doping concentration.  The
sharp quasiparticle peak near the Fermi level is well developed. The peak first
sinks and gradually rises with increasing doping concentration. This behavior
is directly connected with the change of the flat band near the X point with
hole doping.  
\begin{figure}
\includegraphics[width=9.5cm]{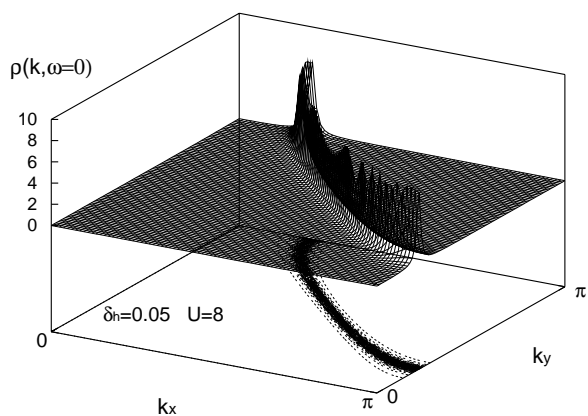}%
\caption{\label{fsn95}
Excitation spectrum at the Fermi energy for $\delta_{h}=0.05$.
}
\end{figure}

\begin{figure}
\includegraphics[width=9.5cm]{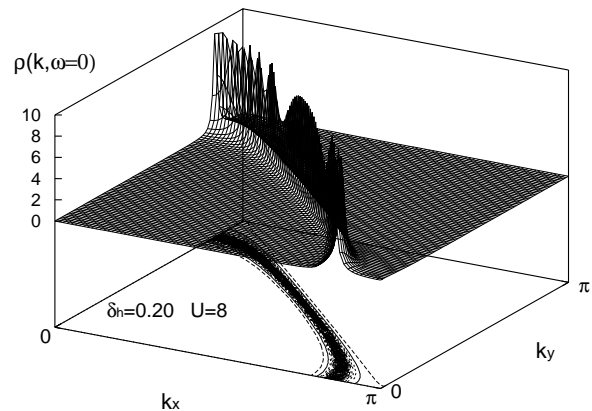}%
\caption{\label{fsn80}
Excitation spectrum at the Fermi energy for $\delta_{h}=0.20$.
}
\end{figure}

We have also investigated the excitation spectrum at $\omega=0$. Results are
shown  in Figs.~\ref{fsn95} and \ref{fsn80}. We find a hole-like Fermi surface
in the underdoped region and an electron-like Fermi surface in the overdoped
region. This behavior is in agreement with the one obtained by the dynamical
cluster approximation at finite temperatures~\cite{maier02}. It should be noted
that the present results do not satisfy Luttinger's theorem~\cite{luttinger60}.
The deviation from the volume predicted by that theorem is small in the overdoped 
region, but it is large in the underdoped region. The ratio of the excess
volume to the predicted one is $\delta v/v_{0} = 0.3$ at $\delta_{h}=0.05$, for
example. It is comparable to the value of 0.4 obtained by the QMC
calculations~\cite{grober00}. 

\begin{figure}
\includegraphics[width=9cm]{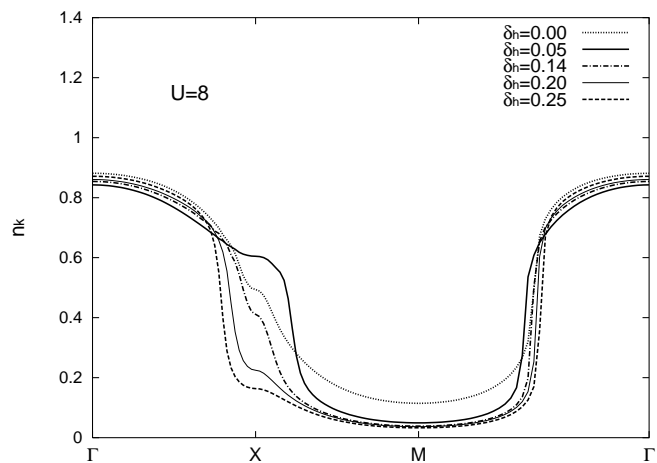}%
\caption{\label{nk}
Momentum distribution along the high symmetry directions for various doping 
concentrations.
}
\end{figure}

The momentum distributions $n_{k}$ along the high symmetry lines are 
presented in Fig.~\ref{nk}. At the M point $n_{k}$ decreases with hole doping
as expected. The momentum distribution at the $\Gamma$ point, on the other
hand, decreases first and then gradually increases with hole doping.
The reduction in the underdoped region originates from the reduction of the LHB
as seen in Fig.~\ref{dosun}. It also yields an enhancement of $n_{k}$ at the X
point. It should be noted that there is no jump of $n_{k}$ at the X point for
half-filling because the quasiparticle weight vanishes there as will be
discussed in the next section. It is also interesting that no discontinuity of
$n_{k}$ is seen at the X point for $\delta_{h}=0.14$ at which the Fermi surface
is not well defined because the flat band is here right at the Fermi
level. These anomalies are not found along the $\Gamma$-M line; we find there  
a clear discontinuity of $n_{k}$ at the Fermi surface irrespective of the
doping concentration. 

\subsection{Marginal Fermi liquid behavior}

\begin{figure}
\includegraphics[width=9.5cm]{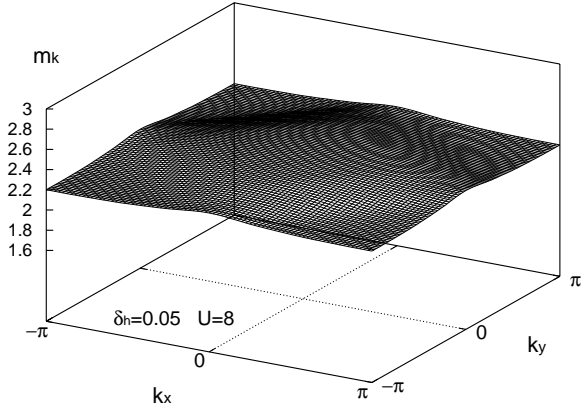}%
\caption{\label{mk95}
Momentum-dependent effective mass for $\delta_{h}=0.05$.
}
\end{figure}

\begin{figure}
\includegraphics[width=9cm]{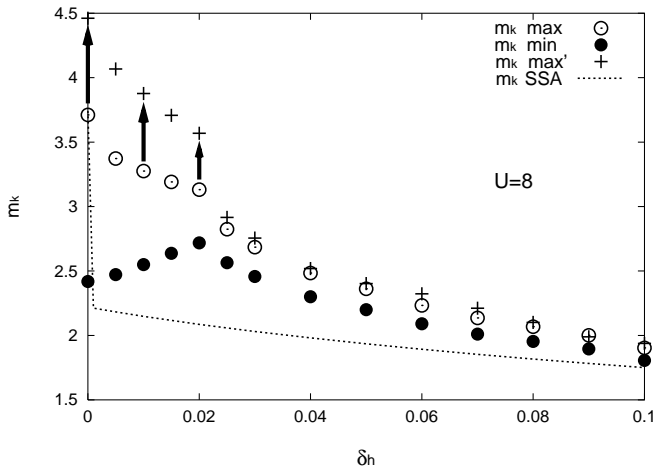}%
\caption{\label{mkn}
Momentum-dependent effective mass $m_{k}$ vs. doping concentration 
$\delta_{h}$. Open circles: maximum value at the X point for
 $\delta_{h} \le 0.02$ and $\Gamma$ point for $\delta_{h} \ge 0.02$,
closed circles: minimum value at the M point. Numerical derivatives are taken
with energy steps of $\delta\omega = 0.05$.  In the case of maximum
$m_{k}$, the results for $\delta\omega = 0.005$ are shown by +. The
momentum-independent effective mass in the single-site approximation (SSA) is
shown by the dashed curve.
}
\end{figure}

We have discussed in Sec. III A the strong $k$ dependence of the effective
mass $m_{k}$ and we have pointed out its logarithmic divergence at the Fermi level
for the case of half filling (see Fig.~\ref{mk1}). When holes are doped, the
effective mass rapidly looses the $k$ dependence. We show in Fig.~\ref{mk95} an
example for 5 \% doping, where the maximum of $m_{k}$ changes from the X point
to the $\Gamma$ point. 
The maximum and minimum values of $m_{k}$ with increasing doping concentration are
shown in Fig.~\ref{mkn}. Both quantities increase with decreasing doping
concentration. At $\delta^{\ast}_{h}=0.025$, we find a sudden change of $m_{k}$.
When the energy in the numerical differentiation is changed in steps varying from 
$\delta\omega=0.05$ to $\delta\omega=0.005$, the maximum value of $m_{k}$ at
the X point increases logarithmically for $\delta_{h} < \delta^{\ast}_{h}$
as shown in Fig.~\ref{mkn}. This implies that up to the finite doping
concentrations of 2.5 \% the electrons are forming a MFL.
\begin{figure}
\includegraphics[width=9cm]{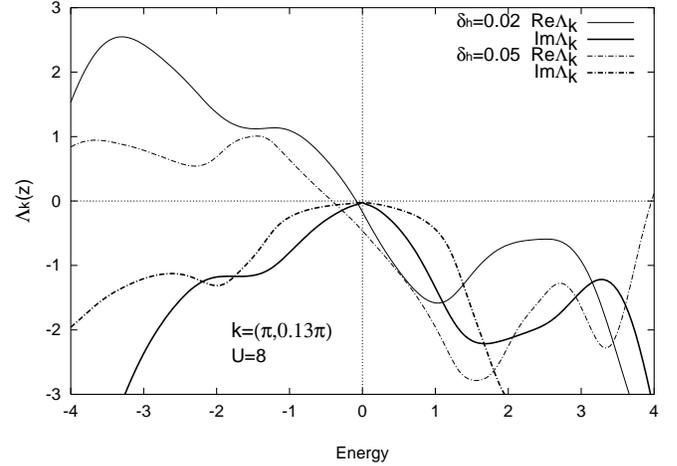}%
\caption{\label{skf}
Self-energy at the Fermi momentum ${\bf k}=(\pi, 0.13\pi)$ 
for doping concentrations $\delta_{h}=0.02$ and $0.05$.
}  
\end{figure}
We can not verify numerically though {\it how accurately} a MFL is realized at
those concentrations. Figure \ref{skf} shows the calculated self-energies at a
particular $\mbox{\boldmath$k$}_{F}$ for $\delta_{h}=0.02$ (in the MFL region)
and $\delta_{h}=0.05$ (in the Fermi liquid region). The Im $\Lambda_{k_{F}}(z)$
at $\delta_{h}=0.02$ is approximately proportional to $|\omega|$ for small
$\omega$. We therefore conclude that the quasiparticle states in the low doping
region are at least very close to a MFL. We note that a MFL can not be
described by the conventional cluster approaches because the long-range
intersite correlations near the Fermi surface are indispensable for it to occur
and they are not taken into account in the usual cluster theories.

\begin{figure}
\includegraphics[width=9cm]{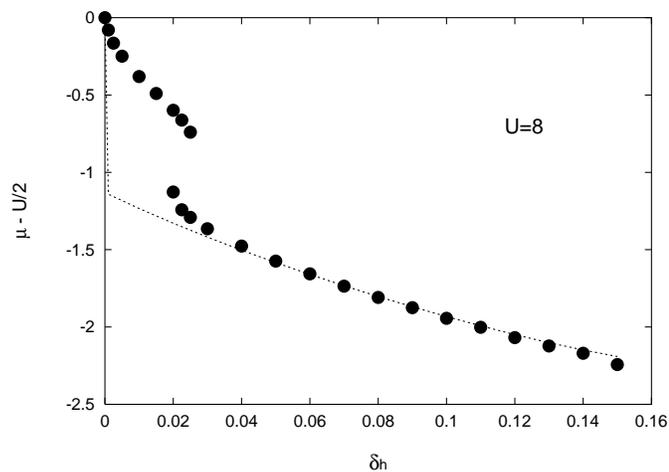}%
\caption{\label{mu}
Chemical potential vs. doping concentration. The dashed curve shows results of
the SSA. 
}
\end{figure}

As mentioned before, the change from the Fermi liquid (FL) to the MFL takes
place discontinuously. This is clearly seen in the doping dependence of the
chemical potential shown in Fig.~\ref{mu}. The chemical potential increases
monotonically with decreasing doping concentration and jumps at
$\delta^{\ast}_{h}=0.025$ by $\Delta\mu \simeq 0.55$. Between $\delta_{h}=0.020$
and $\delta_{h}=0.025$, we find two solutions. Below $\delta_{h}=0.020$, there
is only one solution corresponding to a MFL. 

\begin{figure}
\includegraphics[width=9cm]{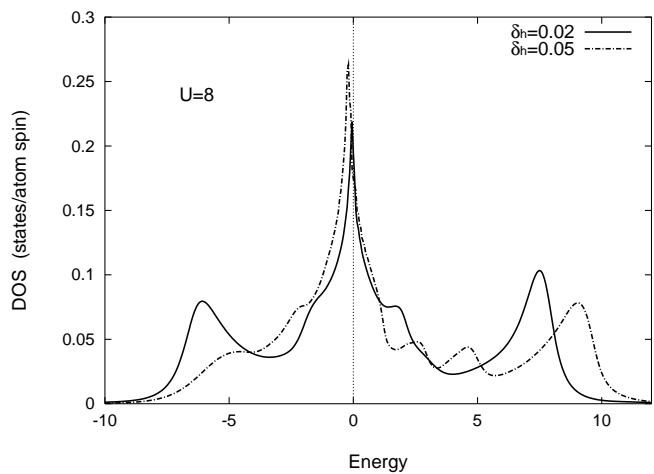}%
\caption{\label{dos98}
The DOS of the marginal Fermi liquid like state (solid curve) and 
the normal Fermi liquid state (dot-dashed curve).
}
\end{figure}

The discontinuity in the chemical potential as well as in $m_{k}$ at
$\delta^{\ast}_{h}$ originates from the collapse of the LHB with increasing hole
doping.  Figure \ref{dos98} shows the total DOS at $\delta_{h}=0.02$ and 0.05.
At low doping concentrations, the spectral weight of the LHB moves directly 
to the UHB with increasing doping because doubly occupied states are 
almost totally suppressed when correlations are strong. Consequently, the van Hove
singularity is pinned to the Fermi level as shown in Fig.~\ref{dos98}.  It
leads to the MFL behavior at finite doping concentrations up to 2.5 \%.
Further doping, however, causes the LHB to disappear and so the peak at the Fermi
level, which is associated with a flat band around the X point, is moving
down below the Fermi level. Finally, the spectral weight of the LHB moves to
the quasiparticle states, and the electrons obtain more and more itinerant
character with reduced correlations.
\begin{figure}
\includegraphics[width=8.5cm]{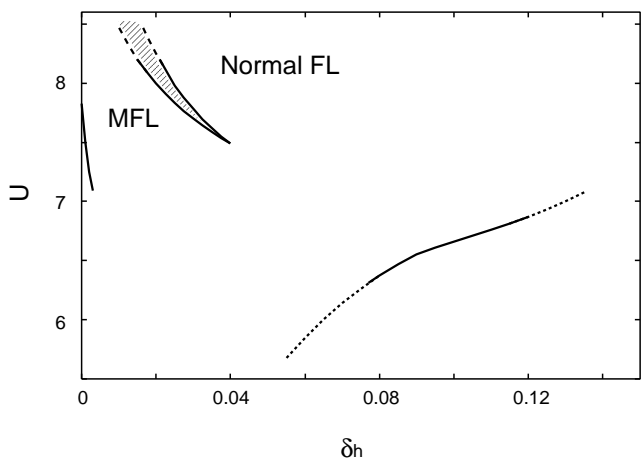}%
\caption{\label{un}
Phase diagram showing the marginal Fermi liquid regime (MFL) 
and the normal Fermi liquid state (Normal FL) regime. The lines at which the
 chemical potential changes discontinuously are shown by solid curves.  The region 
with two self-consistent solutions is shown by the hatched area. Dashed lines along
the hatched area are extrapolations. Along the dotted lines the chemical potential 
has a kink.
}
\end{figure}

Calculated discontinuity lines in the $U-\delta_{h}$ plane are presented in
Fig.~\ref{un}. The MFL-like state with the well-defined LHB is located around
$\delta_{h}=0.02$ and $U=8$. The MFL is separated from the normal FL
state with collapsed LHB by a hatched region in which two self-consistent 
solutions exist. 
Note that the hatched region is slightly reduced as compared with our
preliminary result\cite{kake05-1}
because we have increased the numerical accuracy. 
We also found two additional discontinuity lines in the $U-\delta_{h}$ plane.
One is around $\delta_{h}=0.005$ and $U=7.5$, while the other is around
$\delta_{h}=0.10$ and $U=6.5$. 
The discontinuities are small ($\Delta\mu \sim
0.05$) at both lines and the latter changes to the kink-like 
anomalies which are indicated in
Fig.~\ref{un} by the dotted lines. We have examined these discontinuities of
$\Delta\mu$ by changing the mesh of $\omega$ and $\mbox{\boldmath$k$}$ in the
numerical calculations. 
Our conclusion is that the discontinuity with the hatched region and
that around $(\delta_{h}, U) = (0.005, 7.5)$ exists.  
But, for the line around $(\delta_{h}, U) = (0.10, 6.5)$,
the value $\Delta\mu$ and position seem still to depend on the
number of mesh points even when both the momentum steps and the energy steps
are decreased down to $\pi/80$ and 0.025, respectively. Therefore we do not
exclude the possibility that the line disappears when the numerical
accuracy is increased further. 
\begin{figure}
\includegraphics[width=9cm]{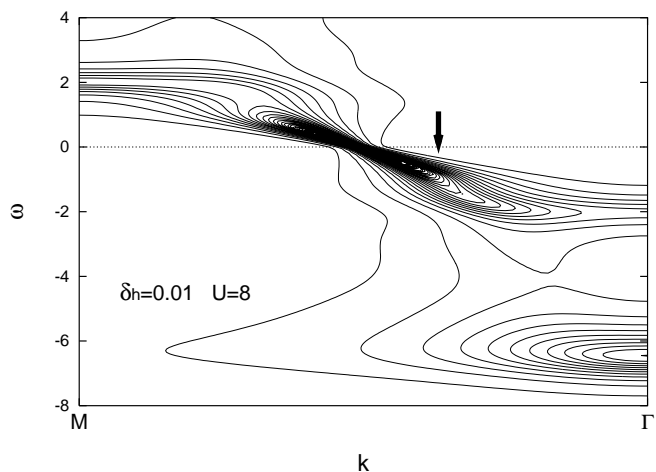}%
\caption{\label{kinkgm}
The excitation spectrum along the nodal direction showing a kink structure for
$\delta_{h}=0.01$. The arrow indicates the kink position 
$|\mbox{\boldmath$k$}|=0.52\pi$.
}
\end{figure}

\subsection{Kink structure in the underdoped region}

We have examined the nonlocal excitations in the MFL region. The excitations in
this region are similar to the ones of the half-filled case (see Fig.~\ref{dosk1}).
The LHB excitations are located around the $\Gamma$ point at $\omega \approx -6.0$.  
We also find flat-band excitations at $\omega = \pm 2.0$, which are due to
short-range antiferromagnetic correlations. The detailed structure of the
quasiparticle states is however unusual.
\begin{figure}
\includegraphics[width=9cm]{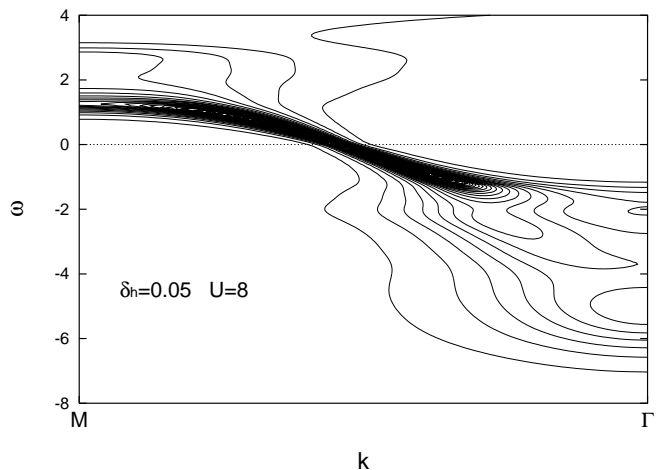}%
\caption{\label{dctr95gm}
The excitation spectrum along the nodal direction for $\delta_{h}=0.05$.
}
\end{figure}
Figure \ref{kinkgm} shows a spectrum along the nodal direction for
$\delta_{h}=0.01$. Notice that a kink structure appears in the quasiparticle states 
at $(|\mbox{\boldmath$k$}|, \omega)=(0.52\pi,-0.85)$. The kink is caused by a
mixing of quasiparticle states with the excitations due to magnetic
short-range order. The latter are enhanced by the nesting of the Fermi surface
in the MFL region.  When $\delta_{h} > 0.03$, the LHB collapses as mentioned
before and the flat band around the X point sinks. Accordingly the excitations
due to antiferromagnetic correlations are weakened.  The quasiparticle state is
then extended to the higher energy region and the kink structure disappears as
shown in Fig.~\ref{dctr95gm}.

\begin{figure}
\includegraphics[width=9cm]{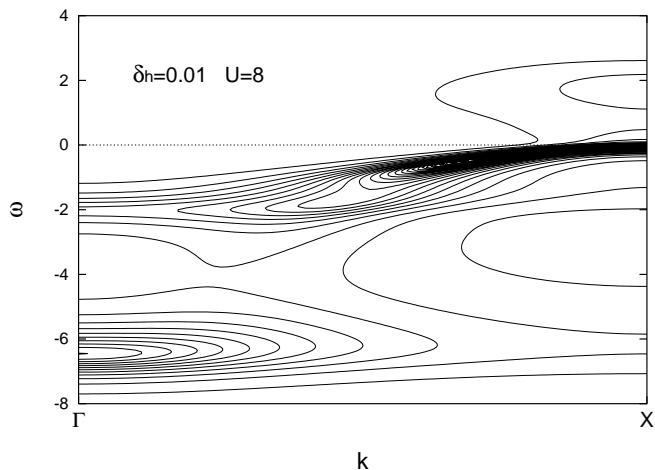}%
\caption{\label{kinkgx}
The excitation spectrum along the antinodal direction for $\delta_{h}=0.01$.
}
\end{figure}

We have also examined the quasiparticle excitations along the antinodal
direction. As seen in Fig.~\ref{kinkgx} the mixing of the quasiparticle
band with the excitations due to the short-range antiferromagnetic correlations
is no longer so strong, and we do not find a clear kink along this direction.

\begin{figure}
\includegraphics[width=9cm]{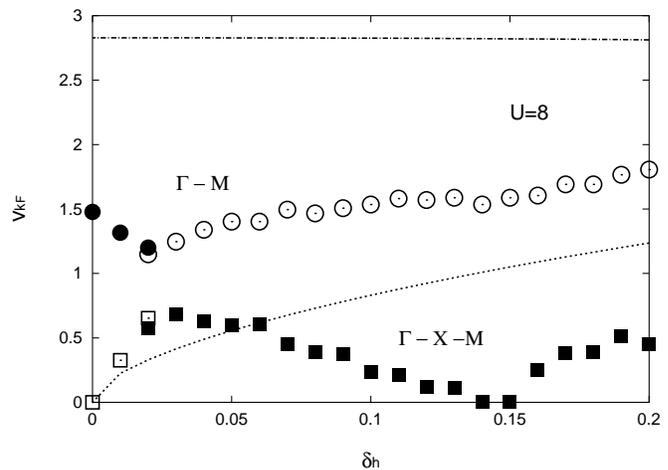}%
\caption{\label{vkf}
Fermi velocity along $\Gamma$-M (closed circles: the MFL state, and open
circles: the FL) and $\Gamma$-X-M (open squares: the MFL state, and closed
squares: the FL). Corresponding values for the noninteracting system are
shown by the dot-dashed curve and dotted curve, respectively.
}
\end{figure}

The ratios of the Fermi velocity above the kink energy $\omega_{\rm kink}$ to
that below the one are $v_{F^{\prime}}/v_{F}=2.7 \ (\delta_{h}=0.0),\ 1.8 \
(\delta_{h}=0.01)$, and $1.5~ (\delta_{h}=0.02)$, respectively. The
concentration dependence of the Fermi velocity above $\omega_{\rm kink}$ is
shown in Fig.~\ref{vkf}. Along the $\Gamma$-M direction, the Fermi velocity is
reduced by a factor of two as compared with the one of the noninteracting system. Its
concentration dependence is rather weak.  The Fermi velocity along the
$\Gamma$-X-M line is strongly influenced by the energy position of the flat
band at the X point. It first increases with the shift of the Fermi momentum 
$\mbox{\boldmath$k$}_{F}$ towards the M point, then decreases gradually with
the return of $\mbox{\boldmath$k$}_{F}$ to the X point. When $\delta_{h}=0.14$, 
$\mbox{\boldmath$k$}_{F}$ crosses the X point, and the Fermi velocity starts to
increase again with hole doping. 

\section{Summary and discussions}

We have investigated numerically the single-particle excitation spectra of the
2D Hubbard model at zero temperature and various doping concentrations, thereby
using the self-consistent projection operator method (SCPM).  The SCPM
describes the nonlocal excitations due to long-range intersite correlations,
and allows us to compute the spectra with high energy and momentum resolution.
The method therefore is able to clarify details of nonlocal excitations
that had not been resolved by previous approaches.   

We have calculated the momentum-dependent excitation spectra from the
underdoped to the overdoped region within the two-site approximation. Pairs of
sites, up to the 50th nearest neighbor were considered. The
spectra are characterized by a LHB around the $\Gamma$ point, and an UHB around the
M point. They show strongly renormalized quasiparticle states near the Fermi
level, as well as excitations caused by magnetic short-range
correlations. We verified that the present results are consistent with those
obtained by QMC calculations at finite
temperatures~\cite{grober00}. Especially the quasiparticle excitations show a
quantitative agreement with the QMC results. We have shown that compared with
the single-site approximation intersite correlations suppress the quasiparticle
weight in the vicinity of the Fermi level 
by a factor of two. They relax also the band narrowing of quasiparticles 
by a factor of two for an intermediate strength of the repulsive 
interaction. Intersite correlations also produce excitations due to
short-range antiferromagnetic order in the underdoped region. Around the X
point the correlations smoothen a sharp drop of the momentum distribution $n_{k}$ at the Fermi surface. 

The antibonding band of Cu 3d$_{x^2-y^2}$ orbitals hybridizing with the
planar O 2p${}_{x}$ and 2p${}_{y}$ orbitals is approximately described by a 2D
Hubbard model~\cite{dagotto94,imada88,norman03}. Therefore the present results
suggest an explanation of some of the unusual properties in the high $T_{\rm
  c}$ cuprates. We obtained a hole-like Fermi surface 
for the underdoped region 
and an electron-like Fermi surface for the overdoped region, being in
agreement with the experimental data on LSCO~\cite{yoshida05}. 
We found numerically a MFL-like behavior at finite doping 
concentrations by analyzing the momentum dependence of 
the effective mass. The results justify a phenomenological
MFL description~\cite{varma89} 
which explains many aspects of the normal state
in cuprates. We found that the MFL is caused by a pinning of the van Hove
singularity to the Fermi level. It is due to a transfer of spectral 
weight from the
LHB to the UHB for small doping concentrations when $U/|t|$ is sufficiently
large. The MFL region calculated here is limited to a small area 
in the $U/|t|$ vs. $\delta_{h}$ plane. 
It remains work for the future to extend the MFL region by
going beyond the two-site approximation and by improving the model.  

The present theory yields a large deviation from the Luttinger
theorem~\cite{luttinger60} in the underdoped region in agreement with 
previous investigations~\cite{grober00,maier02}. In our calculations, the
violation is caused by a sudden reduction of the LHB with increasing doping
concentration or Coulomb interaction strength $U$.  The collapse of the LHB
strongly pulls down the flat band around the X point and increases considerably
the volume below the Fermi surface. Accordingly, the spectral weight of the LHB
moves to a higher energy region. It has not yet been clarified however, under
which general conditions the Luttinger theorem can be violated. Experimentally,
LSCO has been reported to satisfy Luttinger's theorem~\cite{yoshida05}, while
Ca${}_{2-x}$Na${}_{x}$CuO${}_{2}$Cl${}_{2}$ was shown to violate it
strongly~\cite{yoshida05,shen05}. In the latter case, it has been observed that
the LHB approaches the Fermi level with hole doping.  This might correspond to
the collapse of the LHB and the transfer of spectral weight to the higher
energy region found in our calculations. 

We have also found in the underdoped region a kink in the excitation spectrum
along the nodal direction.  The kink is caused by a mixing between the
quasiparticle band and the excitation band caused by short-range antiferromagnetic
correlations. This suggests that the kink found in the cuprate is of electronic
origin; in principle there is no need to include any other (e.g. phononic)
degrees of freedom coupled to electrons in 
order to explain the kink structure. In fact, we obtain kink energies of
$\omega_{\rm kink}=$ 130-170 meV when we adopt a reasonable transfer integral
$|t|=0.2$ eV~\cite{maier02}. They are comparable to the experimental values of
60-70 meV~\cite{cuk}. Calculated ratios of the Fermi velocity $v_{F^{\prime}}/v_{F}$
at $\omega_{\rm kink}$ are between 1.5 and 2.7 in the underdoped region.  These
values should be compared with the experimental ones~\cite{zhou05}, i.e.,
1.3-3.0. We suggest that a strong doping dependence of the kink in
La${}_{2-x}$Sr${}_{x}$CuO${}_{4}$ ($0 < x < 0.07$)~\cite{zhou05} can be
explained by a strong doping dependence of the antiferromagnetic short-range
order as found in the present calculations.  

The 2D Hubbard model considered here is the simplest model for cuprates. It is
leading to a Ne\'{e}l temperature $T_{\rm N}=0$ at half filling.  In order to
describe the layered Cu-based perovskites with high $T_{\rm N}$ we have to
adopt more realistic models, i.e., we have to include 3D features. Furthermore
the present calculations neglect the self-consistency of the off-diagonal
matrix elements of the effective medium.  Because of
the missing self-consistency, the results presented here may underestimate 
the antiferromagnetic correlations. The self-consistency of the off-diagonal
matrix elements of the effective medium and the higher-order cluster
correlations should extend the MFL and the kink region on the phase
diagram. Corresponding calculations are left for future investigations.
\vspace{5mm} \\
{\it Note added in proof}---Recently a high energy anomaly called
``waterfall'' has been found at about 0.3-0.5 eV in the ARPES of 
high-T${}_{c}$ cuprates [ W. Meevasana {\it et. al.},
cond-mat/0612541v1; J. Graf {\it et. al.}, Phys. Rev. Lett., {\bf 98},
(2007) 067004; D.S. Inosov {\it et. al.}, cond-mat/0703223v1 ].  
According to the data on LSCO, the low-energy 
kink anomaly at 60-70 meV strongly depends on the doping
concentration, while the high energy anomaly does not.  We suggest
that the former corresponds to the kink anomaly we found here because of
the strong concentration dependence and that the latter might be a 
phenomenon related to a mixing between the quasiparticle band and 
the incoherent LHB. 

\section*{Acknowledgment}
Numerical calculations have been partly carried out by using the
facilities of the Supercomputer Center, Institute for Solid
State Physics, University of Tokyo.


\begin{thebibliography}{99}
\bibitem{comm1}
We adopted here the self-energy with parameters 
$(\lambda, \lambda^{\prime})=(0,1)$ in eq. (53) of 
ref.\cite{kake04-2},
because we verified in infinite dimensions that this choice of 
parameters yields a reasonable results near the half filling.
\end{thebibliography}
\end{document}